# IR²QSM: Quantitative Susceptibility Mapping via Deep Neural Networks with Iterative Reverse Concatenations and Recurrent Modules

*Min Li, Chen Chen, Zhuang Xiong, Ying Liu, Pengfei Rong, Shanshan Shan, Feng Liu, Hongfu Sun, Yang Gao\**

***Abstract**—Quantitative susceptibility mapping (QSM) is an MRI phase-based post-processing technique to extract the distribution of tissue susceptibilities, demonstrating significant potential in studying neurological diseases. However, the ill-conditioned nature of dipole inversion makes QSM reconstruction from the tissue field prone to noise and artifacts. In this work, we propose a novel deep learning-based IR²QSM method for QSM reconstruction. It is designed by iterating four times of a reverse concatenations and middle recurrent modules enhanced U-net, which could dramatically improve the efficiency of latent feature utilization. Simulated and in vivo experiments were conducted to compare IR²QSM with several traditional algorithms (MEDI and iLSQR) and state-of-the-art deep learning methods (U-net, xQSM, and LPCNN). The results indicated that IR²QSM was able to obtain QSM images with significantly increased accuracy and mitigated artifacts over other methods. Particularly, IR²QSM demonstrated on average the best NRMSE (27.59%) in simulated experiments, which is 15.48%, 7.86%, 17.24%, 9.26%, and 29.13% lower than iLSQR, MEDI, U-net, xQSM, LPCNN, respectively, and led to improved QSM results with fewer artifacts for the in vivo data.*

## I. INTRODUCTION

Quantitative Susceptibility Mapping (QSM) is a post-processing technique to extract the distribution of tissue susceptibilities from the MRI phases [1-4]. QSM has shown significant potential in investigating various neurodegenerative diseases [5-11], alcohol use disorder [12], healthy aging [13-17] and intracranial hemorrhage [18-20]. Furthermore, it has been utilized in studies of brain function by measuring variations in brain oxygen levels [21-23]. However, QSM reconstruction typically involves several non-trivial intermediate processing steps. First, phase unwrapping should be conducted to remove the phase wraps in MRI phases. Then, a background field removal step is carried out to extract the so-called local field maps from the unwrapped phases. Finally, QSM images are reconstructed via dipole inversion from the local field, which is inherently an ill-posed inverse problem.

There have been many efforts devoted to solving the QSM dipole inversion. The COSMOS method (Calculation Of Susceptibility through Multiple Orientation Sampling) [24] has long been considered a gold standard for QSM reconstruction outside the anisotropic white matter regions. However, this method requires at least three repetitive MRI scans at different head orientations, which is very time-consuming and requires the patients to rotate their heads accordingly, making this method impractical for clinical applications. For the single-orientation QSM, TKD [25] mitigates small dipole kernel values by thresholding, but often causes streaking artifacts and estimation errors. Therefore, some other traditional methods, e.g., iLSQR [26], MEDI [27, 28], SFCR [29], STAR-QSM [30], and LN-QSM [31], have been proposed to solve high-quality QSM images from single-orientation local field maps. However, these methods are usually computationally intensive and need manual parameter tuning for different data.

Recently, deep neural networks have become increasingly popular for solving QSM dipole inversion problems. QSMnet [32] was the first to train a U-net to produce COSMOS-like images from single-orientation local field maps, which was then evolved to QSMnet+ [33] by taking advantage of the data augmentation strategy. QSMGAN [34], LPCNN [35], S2Q-Net [36], and MoDL-QSM [37] further improved this training framework (taking *in vivo* acquired local field data as network inputs, and using QSM images reconstructed by traditional methods as labels) by more sophisticated network designs. Alternatively, DeepQSM [38] and xQSM [39] proposed to use

This work was supported by the National Natural Science Foundation of China under Grant No. 62301616 and 62301352, the Natural Science Foundation of Hunan under Grant No. 2024JJ6530, Hunan Provincial Science and Technology Program (NO.2021RC4008), and the HighPerformance Computing Center of Central South University. HS thanks the support from the Australia Research Council (DE20101297 and DP230101628). Correspondence: Yang Gao (yang.gao@csu.edu.cn), Room 429, Information Building, Central South University, Changsha, China.
Min Li, Chen Chen, and Yang Gao are with the School of Computer Science and Engineering, Central South University, Changsha, China.
Zhuang Xiong and Feng Liu are with the School of Electrical Engineering and Computer Science, University of Queensland, Brisbane, Australia.
Yin Liu and Pengfei Rong are with the Department of Radiology, The Third Xiangya Hospital, Central South University, Changsha, China.
Hongfu Sun is with School of Engineering, University of Newcastle, Newcastle, Australia.
Shashan Shan is with State Key Laboratory of Radiation, Medicine and Protection, School for Radiological and Interdisciplinary Sciences (RAD-X), Collaborative Innovation Center of Radiation Medicine of Jiangsu Higher Education Institutions, Soochow University, Suzhou, China.

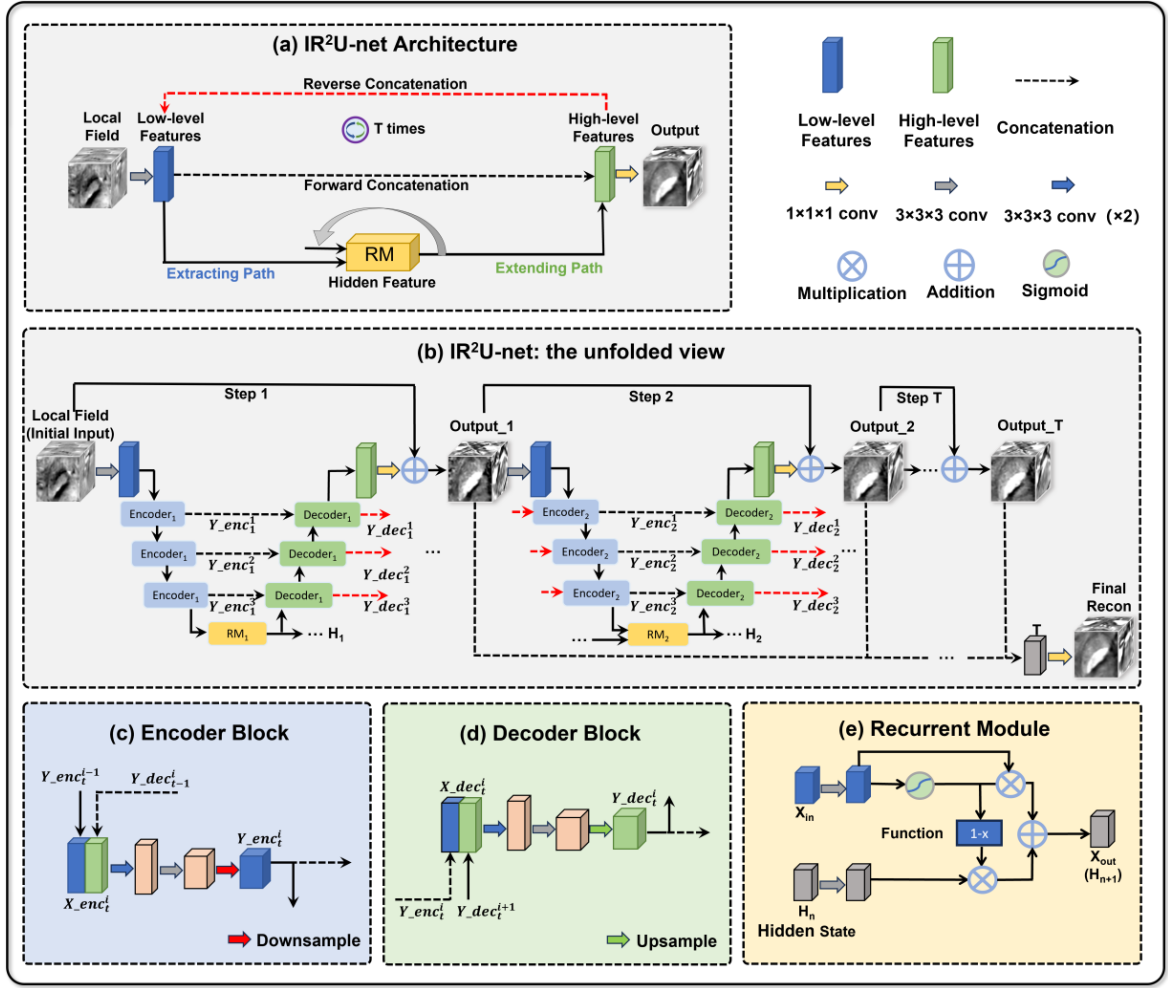

**Fig. 1.** Overview of the proposed IR²QSM trained from the proposed (a) IR²U-net. (b) depicts the detailed and unrolled view of IR²U-net, which is implemented by iterating a specially tailored U-net for *T* times. In addition to the conventional (c) Encoder and (d) Decoder blocks, three Reverse Concatenations (the dashed red line at the top of (a)) and one (e) middle Recurrent Module module are also introduced in the U-net to enhance the efficiency of the latent feature utilization.

pure synthetic or simulated training datasets (generated using the dipole forward model [38]) for deep neural network training, so that their training inputs and labels satisfy the underlying physical model. SAQSM [40] proposed using spatial adaptive modules to reduce susceptibility information loss. AFTER-QSM [41] and MoDIP-QSM [42] could reconstruct QSM images from local field data acquired at different acquisition resolutions and orientations. AdaIN-QSM [43] was capable of resolution-agnostic reconstruction. In addition to these networks for dipole inversion, autoQSM [44] could reconstruct QSM from the total field maps, while iQSM and iQSM+ [45, 46] could extract QSM from the raw wrapped phases in a single step.

Although deep learning techniques have shown promising results in QSM reconstruction, nearly all existing methods were implemented on simple U-nets. In this work, we would like to propose a novel IR²QSM method for dipole inversion via training a specially designed new baseline network, i.e., IR²U-net, on a simulated brain dataset. The proposed IR²U-net is constructed by **I**terating a **R**everse concatenation and **R**ecurrent module enhanced **U-net** four times. Three reverse concatenations from the U-net's extended path to the extracted path were incorporated to fully refine the latent features, and a middle recurrent module was designed for efficient and effective capturing of long-distance information, considering the non-local nature of QSM reconstruction. Comparative experiments on simulated brains, public QSM challenge data, *in vivo* healthy volunteers, and pathological data were conducted to investigate the performances of the proposed IR²QSM with several established QSM dipole inversion methods, including traditional iLSQR, MEDI, and deep learning (DL) based U-net, xQSM, and LPCNN methods.

## II. METHOD

### A. QSM Dipole Inversion

The relationship between the susceptibility $\chi$ and tissue phase $\varphi$ is a convolution with a dipole kernel in the spatial domain, which can be represented by the following equation in k-space [47, 48]:

$$\varphi(\vec{r}) = F^{-1}D(\vec{k}) \cdot F\chi(\vec{r}), \quad (1)$$

where $\vec{r}$ and $\vec{k}$ are the spatial domain and k-space coordinates respectively; $F$ is the Fourier transform; "·" denotes the element-wise multiplication, and $D(\vec{k})$ represents the dipole kernel in the k-space, which can be expressed as the following equation (at pure axial acquisition):

$$D(\vec{k}) = \frac{1}{3} - \frac{k_z^2}{k_x^2 + k_y^2 + k_z^2}, \quad (2)$$

Note that the dipole kernel has a zero-valued double-conical surface ($k_x^2 + k_y^2 = 2k_z^2$) in k-space, making the QSM dipole inversion an ill-posed problem. Equation (1) was also used as the training data generator in many previous DLQSM methods [38, 39, 41].

### B. Network Architecture

#### 1) Overall Structure

The proposed IR$^2$QSM was trained on a novel network backbone, **I**terating **R**everse concatenations and **R**ecurrent modules enhanced **U-net** (IR$^2$U-net), on a simulated dataset. Overall, the proposed IR$^2$U-net is constructed by iterating $T$ times of a specially tailored U-net (Fig. 1(a)), which is designed by inserting reverse concatenation (RC) connections from the expanding to extracting path and a middle recurrent module (RM) into a conventional U-net. The RC and RM designs are beneficial for efficiently processing long-distance information, which helps enhance the non-local QSM dipole inversion.

Figure 1(b) illustrates the unfolded view of the proposed IR$^2$U-net, which can be seen as $T$-cascading U-nets enhanced with the proposed RC and RM modules. As demonstrated in Fig. 1(c) and (d), each encoder or decoder block in the U-net consists of 2 convolutional layers ($3^3$ kernel size with stride 1), 2 batch normalization layers, and 2 rectified linear units (ReLUs), followed by 1 max-pooling layer ($2^3$ kernel size with stride 1) or 1 transposed convolution layer ($2^3$ kernel size with stride 1), respectively. Like conventional U-nets, the upsampling and downsampling data flow enable the network to process multi-scale information by scaling the feature map to different image resolutions. The middle RM module (Fig. 1(e)) is composed of 2 convolutional layers ($3^3$ kernel size with stride 1), 2 batch normalization layers, and 2 ReLUs.

The proposed IR$^2$U-net will produce $T$ latent output images during the $T$ iterative steps. In this work, all these latent outputs are concatenated channel-wisely and fed into a final convolution layer ($1^3$ kernel size with stride 1) to obtain the final output (i.e., the QSM reconstruction), as shown in the center part of Fig. 1(b). In this work, the intermediate U-nets are independent of each other (they are not sharing the same parameters), and we empirically set $T=4$, taking both the performances and computational costs into consideration.

#### 2) Reverse Concatenation

As we have described above, the IR$^2$U-net can be interpreted as $T$ cascading U-nets with three encoders and decoders. Suppose that the input and output feature maps of the $i^{th}$ encoder block in the $t^{th}$ intermediate U-net are denoted by $X\_enc_t^i$ and $Y\_enc_t^i$, respectively, while the features from the corresponding decoder block are denoted by $X\_dec_t^i$ and $Y\_dec_t^i$. Then, the RC connections can be described as follows:

$$X\_enc_t^i = Concat(Y\_enc_t^{i-1}, Y\_dec_{t-1}^i)$$
$$Y\_dec_t^i = \Psi\_dec(X\_dec_t^i), \quad (3)$$

$$Y\_enc_t^i = \Psi\_enc(X\_enc_t^i)$$

where $\Psi\_enc$ and $\Psi\_dec$ denote the cascading operations in the encoder/decoder blocks (2 convolution + 2BN + 2ReLU + max-pooling or transposed convolution). The input to the encoder blocks ($X\_enc_t^i$) is dependent on both the latent features of the current time step ($X\_enc_t^{i-1}$) and the previous time step ($X\_dec_{t-1}^i$), forming a reverse data flow from the high-level features to the low-level features of the proposed backbone, as illustrated in Fig. 1(a).

#### 3) Recurrent Module

As shown in Fig. 1(e), the RM module is inspired by the architecture of Recurrent Neural Networks (RNNs) [49-51]. Suppose that the input of the RM block (the $t^{th}$ intermediate U-net) is $X_t$, and the hidden state from the previous iterations is represented by $H_{t-1}$. Then, the RM can be described as Eq. (4):

$$\alpha = \text{sigmoid}(Conv3D(X_t))$$
$$Y_t = Conv3D(X_t) * \alpha + Conv3D(H_{t-1}) * (1 - \alpha), \quad (4)$$

where $Y_t$ represents the output of the current RM module and is also cached as $H_t$, i.e., the hidden state feature for the $t+1^{th}$ module, and in the first RM module, we set $H_0 = X_1$.

### C. Network Training

All network parameters were initialized using random numbers of a normal distribution with a mean of 0 and a standard deviation of 0.01. All networks were trained for 100 epochs on one Nvidia Tesla A6000 GPU for 60 hours, using the Adam optimizer. The learning rate was set to $10^{-3}$ for the first 30 epochs, $10^{-4}$ for epochs 30 to 60, and $10^{-5}$ for the final 40 epochs; all other hyperparameters were set to their default values. Pretrained IR$^2$QSM networks and related codes can be found at: https://github.com/YangGaoUQ/IR2QSM.

Furthermore, a recently proposed noise-adding module [39] was incorporated during network training to improve the network's performance on the *in vivo* dataset. 5% connections in each layer of the proposed network were also randomly dropped out for better generalization capability, more details about this choice can be found in the Discussion Section (Section V). Note that the noise-adding block and dropout were only added during network training and removed during network inference.

The loss function for the network training in this work consists of two parts:

$$L = \sum_{t=1}^{T} \omega_t MSE(\chi_t, \chi_{GT}) + MSE(\chi_{final}, \chi_{GT}), \quad (5)$$

where $\chi_t$ represents the output of the $t^{th}$ intermediate U-net in Fig. 1(b), $\chi_{final}$ is the final IR$^2$QSM reconstruction, and $\chi_{GT}$ is the training ground truth. $\omega_t = \lambda^{T-t}$ is the weighting parameters between different various loss terms, and we empirically set $\lambda$ as 0.5 in this work.

## III. EXPERIMENTAL SETUP

### A. Training Data Preparation

A total of 14,400 three-dimensional QSM patches (size: $64^3$) were cropped from 96 full-sized QSM volumes (image size: 144×196×128) by traversing the full-sized volumes with a stride of 24×36×20 for network training. The full-sized data were acquired at 3T with 1 mm isotropic resolution and reconstructed using traditional algorithms, i.e., best-path [52] for phase unwrapping, RESHARP [53] for background removal, and iLSQR [26] for dipole inversion. Similar to xQSM [39] and DeepQSM [38], the IR$^2$QSM was trained on simulated datasets generated with Equation (1).

### B. Evaluation Dataset and Comparative Experiments

In this work, the proposed IR$^2$QSM was compared with two iterative algorithms, i.e., iLSQR [26] and MEDI [27], and three deep learning methods, i.e., the U-net, xQSM [39], and LPCNN[35], on both simulated and *in vivo* human brain subjects.

First, 10 brain data were simulated from a publicly available COSMOS label [24] (matrix size: 216×224×160) at 3T with 1 mm isotropic resolution to investigate the effects of the overall network iteration number $T$, RC, and RM on IR$^2$QSM results respectively.

Then, we quantitatively compared the proposed IR$^2$QSM with traditional and deep learning QSM methods using Normalized Root Mean Square Error (NRMSE), High-Frequency Error Norm (HFEN) [8], and Structural Similarity Index (SSIM) [44] on another COSMOS-based simulated data at 3T (1mm isotropic resolution, matrix size: 192×256×176) generated using Eq. (1) and two synthesized data from the 2019 QSM Reconstruction Challenge 2.0 [54] (http://qsm.snu.ac.kr/?pageid=30).

We also quantitatively compared the susceptibility measurements of the proposed IR$^2$QSM with other QSM methods in five deep grey matter regions, including Globus Pallidus (GP), Putamen (PU), Caudate (CN), Substantia Nigra (SN), and Red Nucleus (RN) on a simulated COSMOS brain.

In addition, the performance of IR$^2$QSM was also validated on four in vivo datasets:

(1) A healthy brain data acquired at 3T with the following acquisition parameters: 8 unipolar echoes, matrix size: 256×256×128, 1 mm isotropic resolution, First TE / $\Delta$TE / TR = 3 / 3.3 / 29.8 ms.

(2) A cerebral amyloid angiopathy (CAA) patient was scanned at 3T. The parameters are:10 unipolar echoes, matrix size = 224×224×128, 1 mm isotropic resolution, First TE / $\Delta$TE / TR = 4.2 / 4.1 / 43.4 ms.

(3) A patient with intracranial hemorrhage (ICH) scanned at 3T, with the following parameters: 7 unipolar echoes, matrix size: 180×224×144, 1 mm isotropic resolution, First TE / $\Delta$TE / TR = 4.79 / 4.80 / 40 ms.

(4) A multiple sclerosis (MS) patient scanned at 3T, using the following parameters: 7 unipolar echoes, matrix size: 256×256×128, 1 mm isotropic resolution, First TE / $\Delta$TE / TR = 4.79 / 4.80 / 40 ms.

## IV. RESULTS

### A. The Effects of The Overall Iteration Number T

The influences of the overall iteration number $T$ in the proposed IR$^2$QSM were investigated using 10 simulated brains, and the evaluation metrics were summarized in Table I. Increasing iteration number $T$ from 1 to 4 was effective in improving IR$^2$QSM's reconstruction accuracy, and led to on average 24.89% reduction in NRMSE, 19.43% reduction in HFEN, and 2.58% increase in SSIM. However, it is also found that T=5 did not led to much more improvements, the computational cost went higher as the $T$ increased, and therefore, in this work, we empirically set $T$=4, taking both the network's performances and the computational costs into consideration. Note that in the following sections of the paper,

TABLE I
THE INFLUENCES OF OVERALL ITERATION NUMBER $T$ ON IR$^2$QSM RESULTS

| METHOD | NRMSE(%) | HFEN(%) | SSIM(%) | TIME CPU(S) | GFLOPS |
|---|---|---|---|---|---|
| | MEAN ± STD | | | | |
| $T$=1 | 52.48 ± 0.857 | 39.16 ± 0.450 | 96.23 ± 0.162 | 3.85 | 34.81 |
| $T$=2 | 38.04 ± 0.298 | 30.11 ± 0.361 | 97.35 ± 0.151 | 8.36 | 75.99 |
| $T$=3 | 30.04 ± 0.648 | 21.38 ± 0.262 | 98.58 ± 0.069 | 12.58 | 117.17 |
| $T$=4 | **27.59 ± 0.455** | **19.73 ± 0.346** | 98.81 ± 0.055 | 16.87 | 158.35 |
| $T$=5 | 27.78 ± 0.389 | 19.84 ± 0.522 | **98.82 ± 0.087** | 21.38 | 199.53 |

all the IR$^2$QSM reconstructions, if not specified, are the results of 4 iteration numbers.

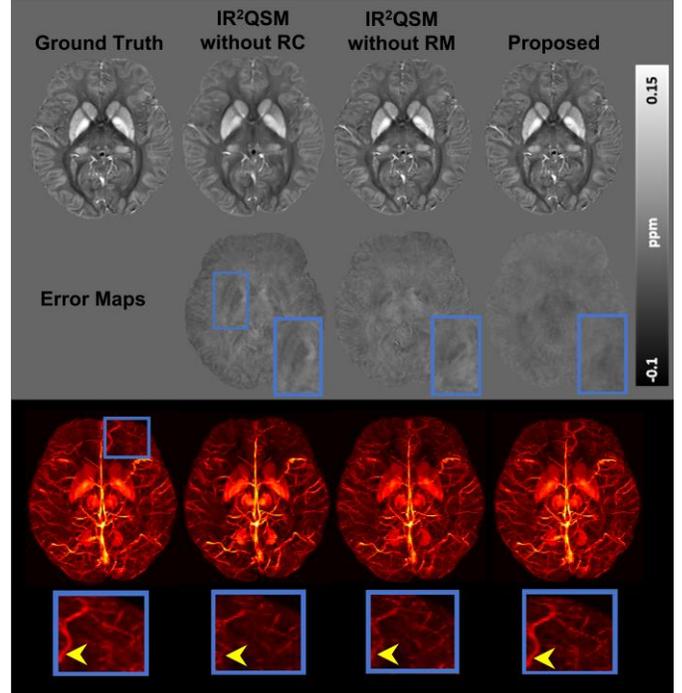

**Fig. 2.** The effectiveness of the proposed recurrent module (RM) and reverse concatenation (RC) on IR$^2$QSM results. The top two rows show the IR$^2$QSM reconstructions with the corresponding error maps relative to the ground truth; the bottom two rows illustrate the maximum Intensity Projection (mIP) results. Yellow arrows point to a blurry vein in IR$^2$QSM without RC or RM.

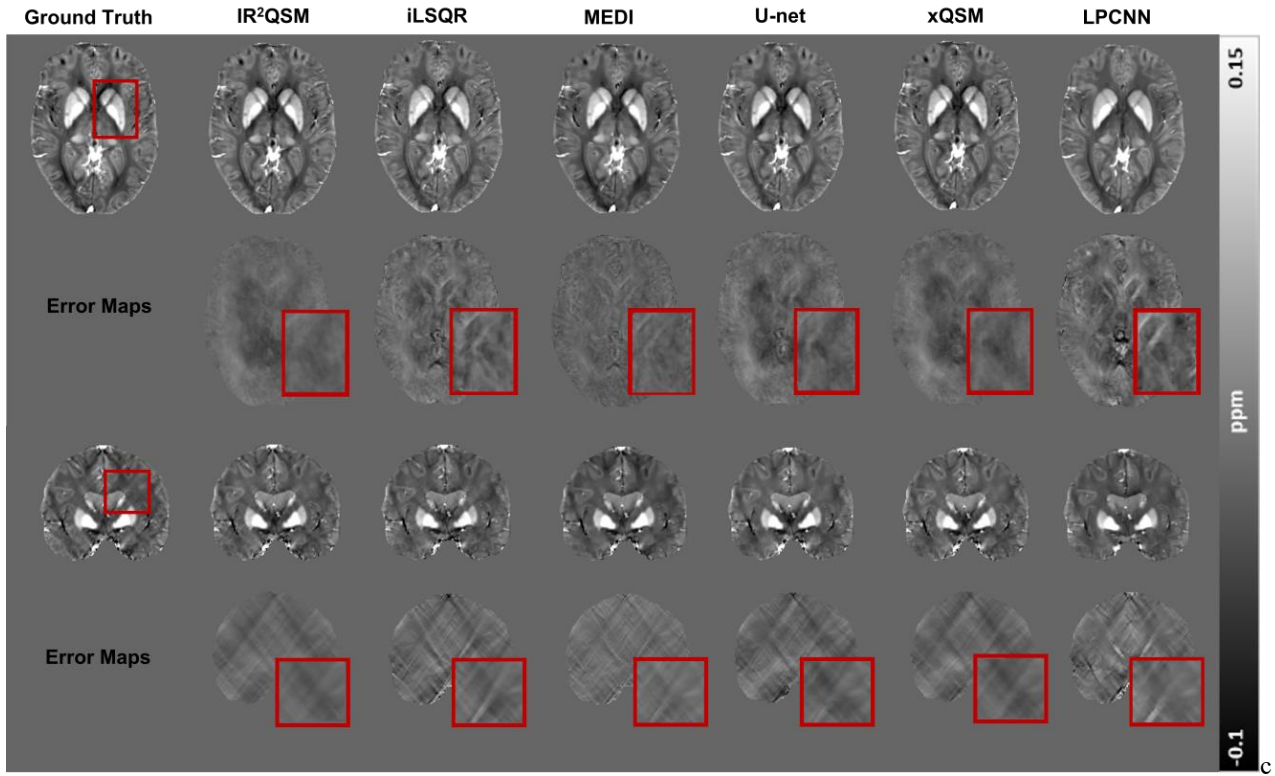

Fig. 3. Comparison of IR$^2$QSM with various QSM methods on a COSMOS-based simulated brain at 3T on two orthogonal planes. Error maps from the zoomed-in regions as highlighted by the red boxes are amplified.

### B. Ablation Study on Recurrent Modules and Reverse Concatenations

An ablation study was also conducted on a simulated brain at 3T to examine the effects of the proposed recurrent module (RM) and reverse concatenation (RC) connections in IR$^2$QSM, and the results are shown in Fig. 2. According to the difference maps and the maximum Intensity Projection (mIP) images, both the RM and the RC designs are effective and could significantly enhance the reconstruction results, as highlighted in the zoom-in error maps and the vessel in mIP (yellow arrows).

The numerical metrics on the simulated brains are reported in Table II. The performances of IR$^2$QSM without RM or RC dramatically dropped, e.g., NRMSE on average increased from 27.59% to 29.56% and 35.81%, which may also implicit that the RC design contributed more than RM (6.25% improvements in NRMSE).

### C. Comparative Studies on Simulation Data
#### 1) COSMOS-Based Simulated Brain Data Results.

Different QSM methods were compared using a COSMOS-based simulated brain (3T, 1 mm isotropic) in Fig. 3. IR$^2$QSM showed the best reconstruction results with visually minimum error maps, and the corresponding numerical metrics (NRMSE, HFEN, and SSIM) were reported in Table III. For instance, the NRMSE of IR$^2$QSM is 27.59%, compared to 43.07%, 35.45%, 44.83%, 36.85%, and 56.72% of iLSQR, MEDI, U-net, xQSM, and LPCNN, respectively.

TABLE II
ABLATION STUDY OF IR$^2$QSM ON THE EFFECTS OF REVERSE CONCATENATION OR RECURRENT MODULE USING TEN SIMULATED BRAINS

| METHOD | NRMSE(%) | HFEN(%) | SSIM(%) |
|---|---|---|---|
| | MEAN ± STD | | |
| IR$^2$QSM without RM | 29.56 ± 0.633 | 21.30 ± 0.345 | 98.54 ± 0.067 |
| IR$^2$QSM without RC | 35.81 ± 0.512 | 26.96 ± 0.201 | 97.83 ± 0.112 |
| IR$^2$QSM (Proposed) | **27.59 ± 0.455** | **19.73 ± 0.346** | **98.81 ± 0.055** |

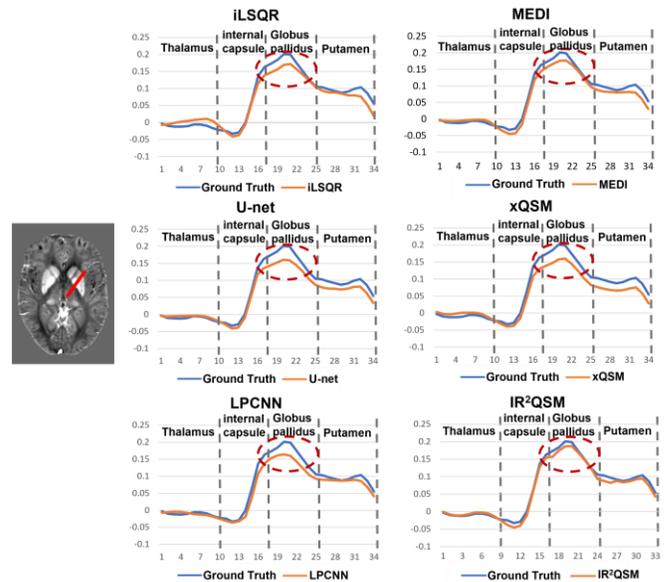

Fig. 4. Susceptibility measurements of different QSM methods on the red line in the deep grey matter regions. Red circles highlight the differences between various QSM results and the ground truth.

TABLE III
COMPARISON OF VARIOUS QSM RESULTS ON A COSMOS-BASED SIMULATED BRAIN

| METHOD | NRMSE(%) | HFEN(%) | SSIM(%) |
|---|---|---|---|
| | | MEAN ± STD | |
| iLSQR | 43.07 ± 1.404 | 40.46 ± 0.665 | 96.21 ± 0.201 |
| MEDI | 35.45 ± 1.133 | 31.18 ± 0.384 | 96.05 ± 0.203 |
| U-net | 44.83 ± 0.641 | 34.72 ± 0.326 | 96.51 ± 0.184 |
| xQSM | 36.85 ± 0.739 | 26.18 ± 0.388 | 97.89 ± 0.113 |
| LPCNN | 56.72 ± 0.563 | 53.42 ± 0.421 | 92.95 ± 0.299 |
| IR$^2$QSM | **27.59 ± 0.455** | **19.73 ± 0.346** | **98.81 ± 0.055** |

Figure 4 compares the susceptibility profiles of different QSM methods along the red line crossing several deep grey matter regions (thalamus (TH), internal capsule (IC), GP, and PU). IR$^2$QSM demonstrated the closest susceptibility measurements compared with the simulation ground truth (highlighted with red circles).

### 2) 2019 QSM Challenge 2.0 Dataset Results

The proposed IR$^2$QSM was compared with various QSM methods using the 2019 QSM challenge 2.0 dataset in Fig. 5. MEDI led to the minimum error maps for contrast "SIM1" (top two rows), while IR$^2$QSM was overall the best DLQSM method and led to the best SSIM for both contrasts and the best HFEN for contrast "SIM2" (bottom two rows) among all algorithms, as confirmed by Table IV.

TABLE IV
COMPARISON OF VARIOUS QSM METHODS ON 2019 QSM CHALLENGE 2.0 DATASET.

| METHOD | NRMSE(%) | HFEN(%) | SSIM |
|---|---|---|---|
| | | SIM1/SIM2 | |
| iLSQR | 69.34/60.27 | 52.31/53.28 | 0.918/0.982 |
| MEDI | **53.61/50.69** | **39.26**/49.47 | 0.960/0.991 |
| U-net | 68.13/67.67 | 47.20/52.04 | 0.970/0.990 |
| xQSM | 66.75/62.95 | 43.07/47.30 | 0.973/0.990 |
| LPCNN | 56.50/56.98 | 52.23/57.53 | 0.898/0.990 |
| IR$^2$QSM | 64.48/58.23 | 39.57/**38.80** | **0.981/0.991** |

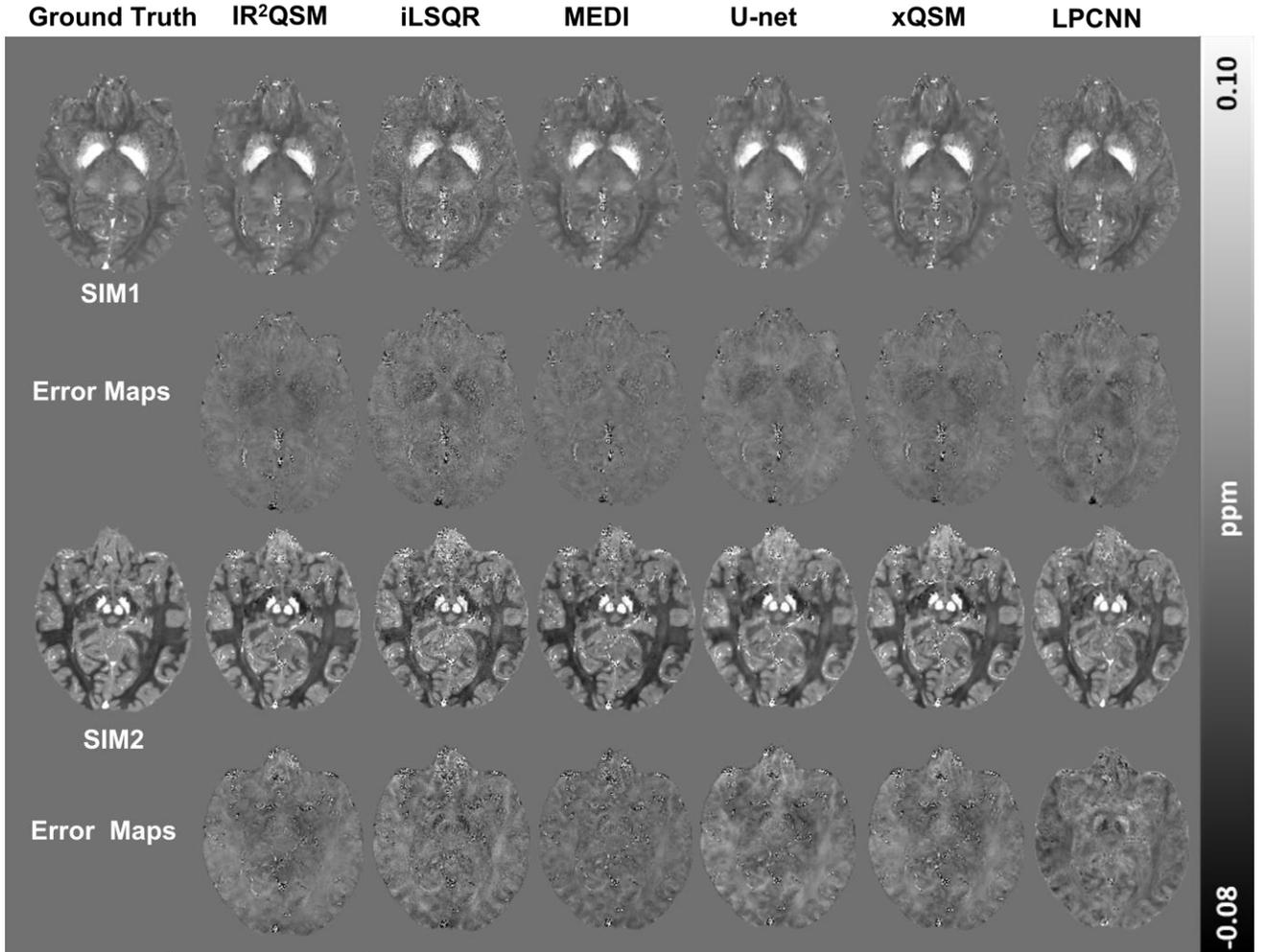

**Fig. 5. Comparison of different dipole inversion methods on the 2019 QSM Challenge 2.0 data. The top two rows show the QSM images and corresponding error maps for data "SIM1", while the bottom two rows depict the results and error maps for data "SIM2".**

### 3) Susceptibility Measurements Analysis in Five DGM Regions

Figure 6 compares susceptibility measurements of different QSM methods in five DGM regions using linear regression. IR$^2$QSM achieved the highest accuracy relative to the ground truth in GP, CN, and RN regions with fitting slopes of 0.94, 0.95, and 0.97, respectively. In particular, in the CN region, the fitting slope of IR$^2$QSM is 0.95, and that of iLSQR, MEDI, UNet, xQSM, and LPCNN is 0.87, 0.91, 0.85, 0.91, 0.87, respectively. In addition, IR$^2$QSM also demonstrated minimum mean squared errors (MSE) in GP, PU, and CN regions and the highest coefficient of determination ($R^2$) in PU and CN regions, as detailed in Table V.

TABLE V
QUANTITATIVE ANALYSIS OF SUSCEPTIBILITY MEASUREMENTS RESULTS IN FIVE DGM REGIONS.

| METHOD | GP | PU | CN | RN | SN |
|---|---|---|---|---|---|
| | MSE($10^{-5}$)/$R^2$ | | | | |
| iLSQR | 19.90/0.880 | 8.17/0.967 | 7.48/0.960 | 14.67/0.948 | 19.36/0.890 |
| MEDI | 4.44/**0.990** | 2.99/0.995 | 3.74/0.988 | **11.87**/0.972 | 5.969/**0.992** |
| U-net | 10.92/0.963 | 4.80/0.989 | 8.32/0.975 | 22.37/0.885 | 17.93/0.890 |
| xQSM | 11.36/0.942 | 4.26/0.985 | 4.08/0.986 | 15.89/0.893 | **4.78**/0.986 |
| LPCNN | 65.79/0.784 | 28.72/0.914 | 22.06/0.852 | 162.8/0.105 | 27.46/0.977 |
| IR$^2$QSM | **4.43**/0.979 | **2.61**/**0.995** | **2.86**/**0.991** | 12.21/0.914 | 5.19/0.963 |

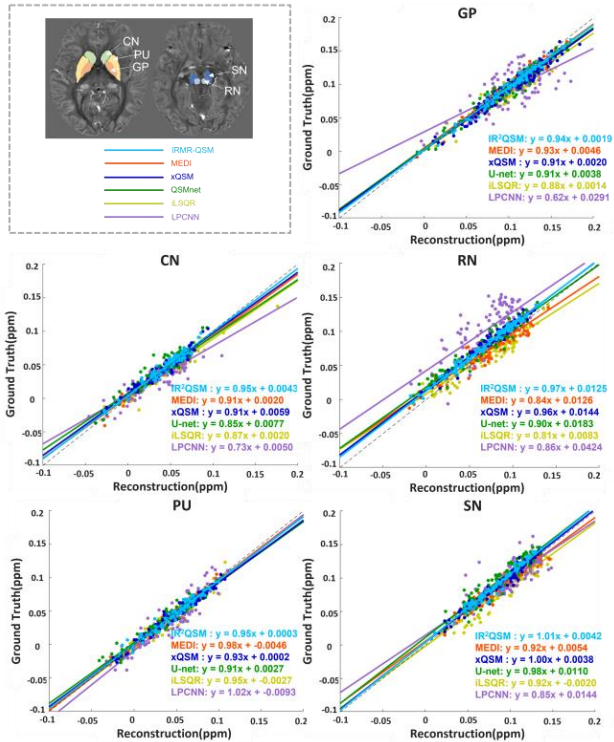

**Fig. 6.** Scatter plots of susceptibility measurements in five deep grey matter regions of various QSM methods (x-axis) against the ground truth (y-axis). Linear regression results are reported correspondingly.

### D. In Vivo Healthy Subjects

Figure 7 compares IR$^2$QSM with other QSM methods on an *in vivo* brain acquired at 3T. IR$^2$QSM showed comparable results with COSMOS and other existing QSM dipole inversion methods in deep grey matter regions, as highlighted in the zoomed-in images in Fig. 7(a). The susceptibility measurements of five DGM were also quantitatively compared in Fig. 7(b). Statistical *t*-tests (carried out based on all voxels inside the ROIs) found that IR$^2$QSM led to similar quantitative measurements in GP compared to COSMOS, while iLSQR, MEDI, and U-net exhibited significantly underestimated results, on average 0.024 (*P*=0.003), 0.019 (*P*=0.015), and 0.012 (*P*=0.028) ppm lower, respectively.

### E. In Vivo Pathological Brain Data

Figure 8 compares the proposed IR$^2$QSM results with iLSQR, MEDI, U-net, xQSM, and LPCNN using three *in vivo* pathological subjects. All methods successfully detected the pathological lesions in all three brains; however, iLSQR, MEDI, and LPCNN led to more artifacts in the sinus region, as highlighted by the yellow arrows. In addition, LPCNN also presented over-smoothing results with suppressed contrasts.

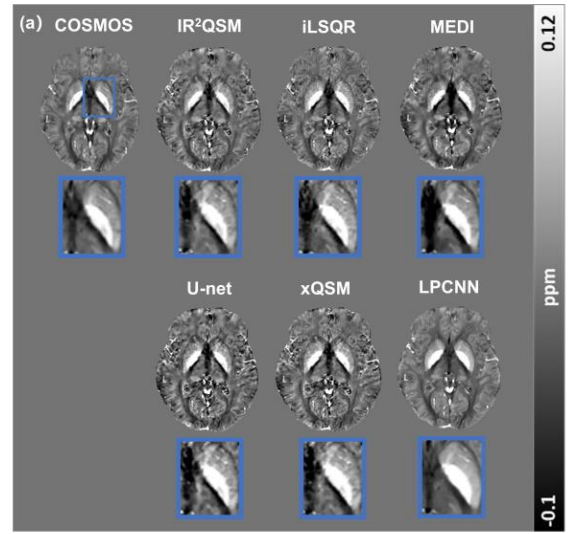

**Fig. 7.** Comparison of various QSM methods on an *in vivo* subject acquired at 3T. (a) shows the QSM results of different methods, and the bar graphs in (b) compare the susceptibility measurements of IR$^2$QSM with other QSM methods in five DGM regions.

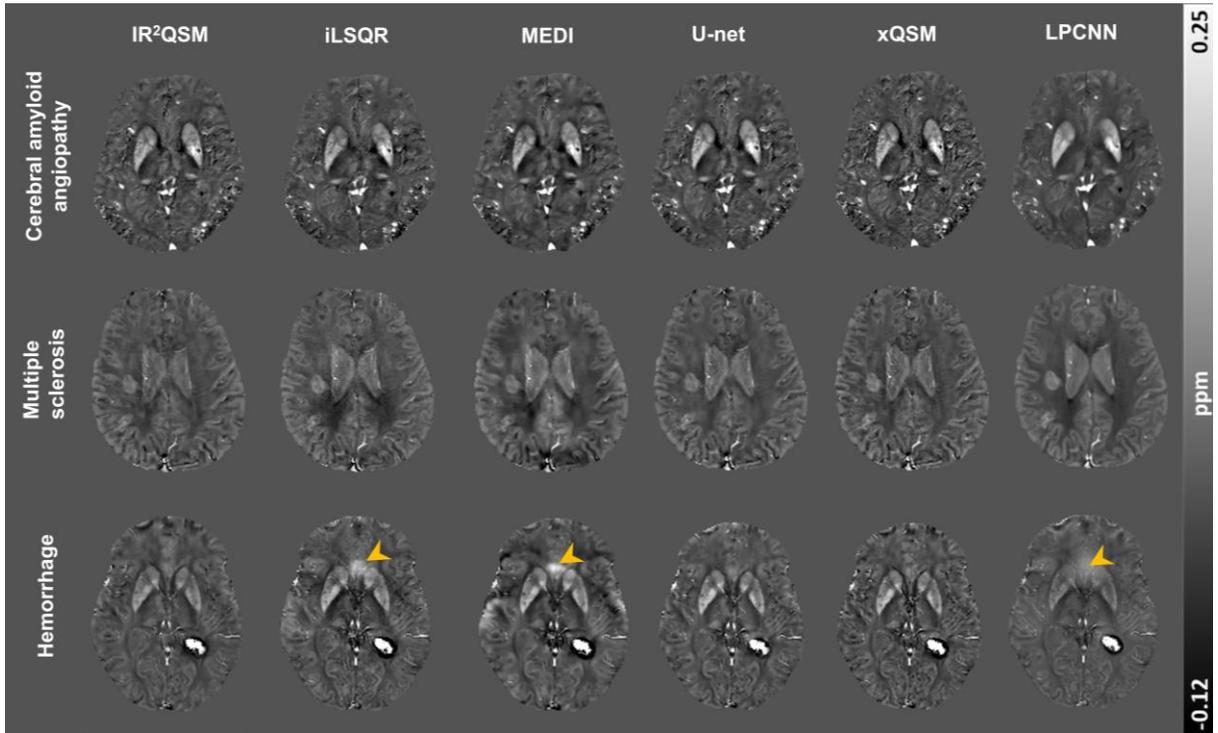

**Fig. 8.** Comparison of IR$^2$QSM with various QSM methods on three *in vivo* pathological brain data from patients with cerebral amyloid angiopathy, multiple sclerosis, and intracranial hemorrhage, respectively. Yellow arrows point to the artifacts in iLSQR, MEDI, and LPCNN methods.

## V. Discussion

In this study, we proposed a new DLQSM dipole inversion method, namely IR$^2$QSM, which was trained on our proposed IR$^2$U-net on a simulated training dataset. The novel IR$^2$U-net was constructed by iterating four times of a specially tailored U-net, which is empowered by Reverse Concatenations and Recurrent Modules. IR$^2$QSM was compared with two traditional and several state-of-the-art deep learning methods (i.e., iLSQR, MEDI, U-net, xQSM, LPCNN) using a comprehensive brain dataset, including both simulated and in vivo brains. It showed the best numerical performances (e.g., NRMSE, HFEN, and SSIM) on the simulated brains and also presented QSM images with the least noise and artifacts on the in vivo dataset, and in the meantime, it successfully alleviated the over-smoothing and susceptibility underestimation in LPCNN results.

The key designs in the proposed IR$^2$QSM are the Reverse Concatenations (RC) and Recurrent Modules (RM). With the help of RCs from the U-net's expanding path to the extracting path, the network could efficiently merge high-level semantic features with low-level semantic features. Meanwhile, the RM facilitated long-distance information processing with an RNN-like recursive design. Overall, these two modules resulted in dramatically improved numerical metrics, as demonstrated in the ablation studies.

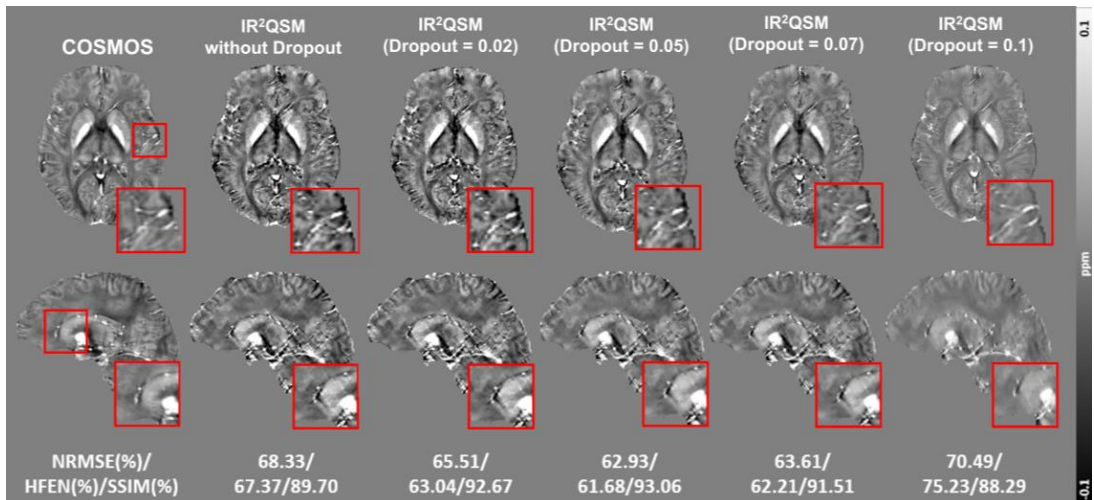

**Fig. 9.** The influences of dropout rate during network training on IR$^2$QSM results on a healthy subject acquired at 3T. Quantitative numerical metrics between the reconstructed images and a COSMOS reference are also reported correspondingly.

As reported in a previous review paper [55], DLQSM methods trained on simulated or synthesized datasets are more prone to noise and artifacts in the *in vivo* subjects. To alleviate this problem in IR$^2$QSM, in addition to our previously proposed noise-adding layer in xQSM, we also adopted the dropout technique [56, 57] during network training to improve IR$^2$QSM generalization capability. Figure 9 compares IR$^2$QSM trained without and with dropout (at 0.02, 0.05, 0.07, 0.1 rates) on an *in vivo* data acquired at 3T. The results indicated that dropping out an appropriate proportion (i.e., 0.02, 0.05, and 0.07) of the network connections during network training was very effective in mitigating the noise and artifacts in the *in vivo* data, compared with IR$^2$QSM results without any dropout. However, relatively higher dropout rates (0.1) could lead to significant over-smoothing in the QSM results, and in this work, we manually set a 0.05 rate as our final choice, which also led to the most similar results to the COSMOS reference, with the best NRMSE/HFEN/SSIM metrics.

In this work, although IR$^2$QSM showed improved results compared with previous DLQSM methods, i.e., U-net, xQSM, and LPCNN, the computational loads (e.g., the FLOPs and reconstruction speeds) of IR$^2$QSM also dramatically increased, which could be a drawback in speed-demanding applications. Additionally, we limited our experiments to human brain data acquired with 1 mm isotropic resolution (consistent with the training data) only, and in the future, more comprehensive tests of different resolutions or subjects beyond human brains should be carried out to further validate the proposed method. Future work will also investigate more advanced network backbones for faster and more precise DLQSM reconstruction in the meantime.

## VI. CONCLUSION

In this work, we proposed a novel IR$^2$QSM method for QSM dipole inversion and conducted comparative experiments based on a comprehensive dataset including both simulated brains and *in vivo* subjects. The results demonstrated that IR$^2$QSM led to results with better numerical metrics and improved robustness to noise and artifacts, compared with multiple established traditional and deep learning methods.